# Mixed Ion Beams Enable Simultaneous Treatment and Real-Time Imaging in Carbon Ion Therapy

Lennart Volz[1,†], Ronja Hetzel[1,†], Maximilian Dick[1], Maria Chiara Martire[1,6], Guangru Li[1], Christoph Schuy[1], Sali Ballouz[1], Mikaël Simard[2], Saad Shaikh[2], Charles-Antoine Collins-Fekete[2], Tim Wagner[1], Michael Galonska[1], Andrii Patushenko[1], Ralph Hollinger[1], Fabio Maimone[1], Jens Stadlmann[1], Lars Bozyk[1], David Ondreka[1], Simone Savazzi[3], Marco Pullia[3], Marco Durante[1,4,5] and Christian Graeff[1,6,*]

[1]GSI Helmholtzzentrum für Schwerionenforschung, Darmstadt, Germany
[2]Department of Medical Physics & Biomedical Engineering, University College London, London, UK
[3] Centro Nazionale di Adroterapia Oncologica (CNAO), Pavia, Italy
[4]Department of Condensed Matter Physics, Technische Universität Darmstadt, Darmstadt, Germany
[5]Department of Physics "Ettore Pancini", University Federico II, Naples, Italy
[6]Department of Electrical Engineering and Information Technology (ETIT), Technische Universität Darmstadt, Darmstadt, Germany
[†] These authors contributed equally to this work, and share first authorship
* Correspondence: c.graeff@gsi.de

**ABSTRACT**. Carbon ion therapy is one of the most advanced forms of radiotherapy, promising improved efficacy against resistant cancers. However, the high precision offered by the carbon ion Bragg peak requires precise knowledge of the beam range inside the patient. We report the first experimental realization of range monitoring and portal imaging with a mixed ion beam, where carbon ions are treating the tumor while helium ions simultaneously accelerated to the same velocity fully traverse the patient and provide treatment feedback. Using the GSI synchrotron, a beam of $^{12}C^{3+}$ and $^{4}He^{1+}$ ions is accelerated, exploiting their nearly identical charge-to-mass ratios. Stable extraction with controlled helium fractions down to 7% is demonstrated. Beam characterization reveals that the helium ion Bragg peak can be cleanly separated from the carbon ion fragment background which enables accurate detection of sub-millimeter Bragg peak displacements. Mixed-beam radiographs of a lung-cancer-like phantom offer target position detection to better than 0.5 mm accuracy. This establishes mixed beams as a powerful modality for real-time image guidance in carbon ion therapy, uniquely providing simultaneous treatment delivery, range probing, and portal imaging. By overcoming range uncertainty inside the patient, mixed beams will enable to fully exploit the precision of carbon ion therapy.

## I. INTRODUCTION.

Carbon ion radiotherapy (CIRT) represents one of the most advanced forms of cancer treatment, combining highly localized dose deposition with superior biological effectiveness [1]. Its key physical advantage lies in the Bragg peak, which enables conformal dose delivery to deep-seated tumors while sparing surrounding tissue [2]. However, this precision is undermined by range uncertainties inside the patient, which are exacerbated by anatomical motion and conversion errors in CT-based stopping-power prediction [3]. To maintain treatment safety, large margins are currently applied, blunting the very advantage of heavy-ion therapy and limiting its potential in challenging sites such as lung, liver, or pancreas [4-6].

Numerous strategies have been explored to address this problem [7,8]. Techniques based on detecting secondary radiation, such as positron emission tomography (PET) [9,10] or prompt gamma imaging [11,12], have provided valuable insights, but remain limited by low signal statistics or insufficient real-time capability for carbon ion beams. To increase the PET signal it would be possible to use radioactive ion beams [13], but this technique is still far from clinical implementation. Anatomical imaging approaches, including online fluoroscopy [14] and magnetic resonance imaging (MRI) [15], promise real-time guidance but require indirect range inference and involve significant technical or dosimetric compromises. The ideal guidance would provide both real-time actionable range and anatomy information at little or no extra dose to the patient. This central challenge of ion therapy—reliable, real-time knowledge of the beam range in the patient—remains unresolved.

A fundamentally different strategy has been proposed: the use of mixed ion beams, in which a therapeutic ion species is co-accelerated with a lighter ion of similar magnetic rigidity [16-20]. In particular, helium ions have three times the range of carbon ions at equal velocity, making them ideally suited as an in-beam probe for portal imaging and range verification [16,17]. The concept is shown schematically in Figure 1. Recently, we have demonstrated for the first time the production of a simultaneously accelerated mixed carbon–helium beam at the GSI synchrotron [19]. A mixed beam has now also been achieved at a clinical facility [20].

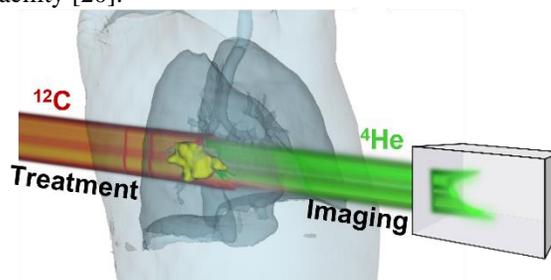

*Figure 1:* Schematic representation of a mixed beam simultaneous treatment and monitoring. The therapeutic carbon ion beam (red) stops in the tumor (yellow) to deliver the treatment, while a small percentage of helium ions (green) mixed into the beam crosses the patient. It can then be monitored in a suitable detector for real-time treatment feedback. The patient body is visualized in grey, the lungs in dark grey.

Previous studies demonstrated either mixed-beam production, or imaging concepts using separate beams; here we demonstrate, for the first time, simultaneous therapeutic irradiation and real-time imaging using a single, clinically relevant mixed carbon–helium ion beam at the GSI synchrotron. We show stable beam production, quantify the dosimetric properties, and show proof-of-concept experiments for concurrent dose delivery, range verification, and portal imaging. The characterized mixed beam was used in simulations on patient CT data, to assess its clinical potential This demonstration establishes mixed beams as a viable pathway towards real-time image guidance in heavy-ion therapy, addressing one of its most critical limitations and opening new directions for both clinical and fundamental applications. Experimentally, we first characterize beam stability and dosimetry (Section II.B), then demonstrate concurrent range monitoring (Section II.C), followed by mixed-beam portal imaging (Section II.D). Finally, we assess clinical relevance through patient-specific simulations and deep learning–based range inference (Section II.E).

*Contact author: C.Graeff@gsi.de

## II. METHODS

### A. Mixed beam production

Ions were extracted from GSI's Electron Cyclotron Resonance (ECR) ion source, using methane with helium as a support gas. The source and low-energy beam transport were tuned to A/Q=4, with the aim of producing approximately 130 µA of $^{12}C^{3+}$ and a variable current of $^{4}He^{1+}$. The source was operated for 14 days prior to beam time to achieve clean, stable working conditions. Different intensities of helium ions in the beam were investigated by changing the concentration of the helium support gas. Measurements were performed with a 'low', 'medium', and 'high' helium concentration setting, with exact ratios determined post-hoc. The 'low' setting corresponded to a goal of ≤10% helium ions in the mixed beam, as was used in previous simulation studies and experiments with separate beams [17]. The 'medium' helium concentration setting was aimed at yielding a ~20% helium ion concentration, which would be the approximate maximum regarding the additional dose in the treatment (≤1% of the RBE weighted target dose). For mixed beam imaging, a 'high' helium concentration was used, where the ratio was increased to about 1:1 between carbon and helium ions in the beam. In all experiments, a 5% to 7% oxygen contamination was present, likely originating from the ECR source walls. Strategies for eliminating this contamination are discussed in Section IV.

The mixed beam was accelerated in the linear accelerator UNILAC to an energy of 11.4 MeV/u, injected without further stripping into the SIS18, where the beam was accelerated to either 225 MeV/u and 260 MeV/u. Slow extraction was then performed either by transversal radio-frequency knock-out (RFKO). During the acceleration in UNILAC and SIS18, helium and carbon ions behave essentially identical. Slow extraction can, however, be tuned to be more or less sensitive to the small mass difference between the two ion species by varying the horizontal chromaticity of the synchrotron. For RFKO, chromaticity was reduced from the natural chromaticity of about -6 to a value close to zero, resulting in the reasonably flat helium to carbon ratio over the spill. This setting was utilized to carry out the actual physics experiments. In addition, a small accelerator study was performed to investigate the effect of varying horizontal chromaticity on the behavior of carbon and helium over the spill in tune-sweep extraction. Depending on the value of horizontal chromaticity, the helium component can in fact be shifted in time with respect to the carbon component, the extremes being perfect overlap or complete separation of the two components [21].

For the 'high' helium concentration, the SIS18 spill-optimization system was used, in order to obtain a reasonably flat beam intensity over the spill. Diagnostic elements capable of separating the ion species in the mixed beam are available only in the low energy beam transport (LEBT) section behind the ion source, where a mass spectrometer separated ions by their A/Q ratio with a given energy per nucleon. While the spectrometer is thus not able to distinguish between the 12C3+ and 4He+ content in the mixed beam, when subtracting spectra with and without He, one can see the 4He+ fraction indirectly in terms of current. In addition, a rough estimate of the helium to carbon ratio is obtained by recording the optical emission spectrum of the ion source plasma with a set-up comprising a telephoto lens and an optical spectrometer (OceanOptics QE Pro), evaluating emission lines of carbon (wavelength 465 nm) and helium I (728 nm) [22].

Throughout the acceleration and beam transport up to GSI's medical research room (Cave M) [23], none of the available beam diagnostic elements can distinguish ion types, since the integral measurements are dominated by carbon ions, which completely eclipse possible helium ion signals.

### B. Beam characterization

First, the mixed beam was characterized to assess its potential for concurrent treatment and verification at Cave M. Of importance for medical application is the integrated depth-dose distribution (IDD), as well as the temporal and lateral alignment of the mixed beam components.

#### 1. Ionization chamber setup

The IDDs and temporal alignment of the mixed beam was analyzed using a setup of three parallel plate ionization chambers (IC1-3) in between which two variable range shifters (RS1 and RS2) were mounted. Between RS2 and IC3, three 50 mm blocks of PMMA were added to account for the factor 3 longer helium range. The setup is shown in Figure 2.

The ICs had an active area of 22 cm x 22 cm, with a gap of 20 mm, and were operated at 1800 V. All ICs were gassed with an 80:20 mix of Ar:CO2. Their signals were amplified (SR570, SRS, Sunnyvale, CA, USA), with all three signals digitized to 14 bits in an FPGA scope card (PXIe-5172, NI, Austin, TX, USA) at a frequency of 50 kHz. The variable range shifters hosted a set of 8 (RS1) and 10 (RS2) PE plates, with a water-equivalent thickness (WET) of the first plate of

*Contact author: C.Graeff@gsi.de

0.5 mm (RS1) and 0.06 mm (RS2), respectively. The following plates approximately double in WET in each step, for a total possible range shift of 127.3 mm (RS1) and 61.6 mm (RS2) at a resolution of the respective thinnest plate.

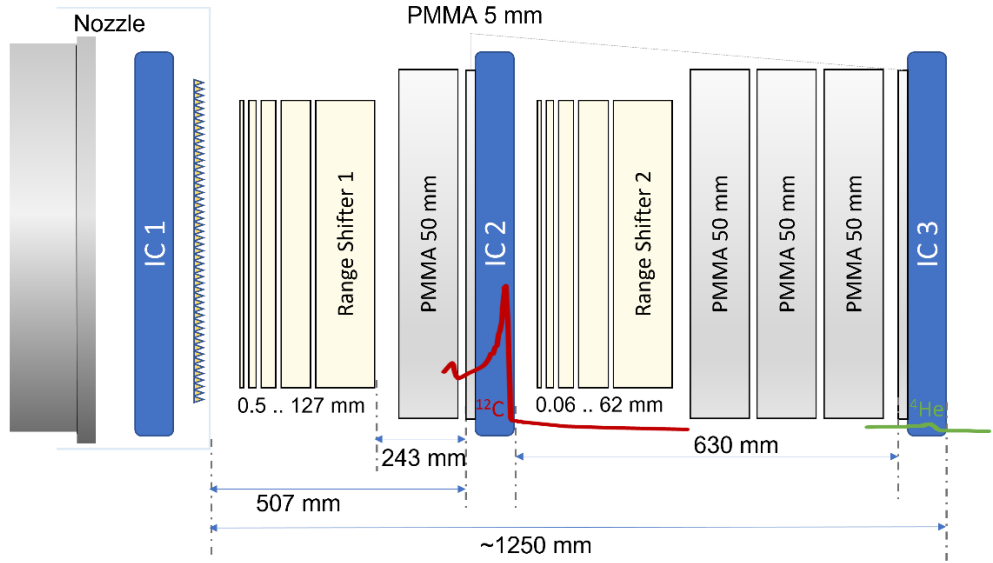

*Figure 2*: Schematic of the experimental setup, consisting of three Ionization Chambers (IC1-3), separated by PMMA bolus elements and two range shifters, which are used to step through the water-equivalent range to both IC2 and IC3. In this way, depth dose profiles of carbon and helium ions can be separately measured at the same time.

IC1 served to normalize the signals of IC2 and IC3, and was mounted directly in the nozzle. With RS1 and RS2, the WET leading up to IC2 and IC3 could be adjusted, enabling to capture the IDD of the full beam, otherwise not possible with commercial equipment. In addition, this setup allowed to simultaneously position the carbon ion Bragg peak in IC2 and the helium ion Bragg peak in IC3, by adjusting RS1 and RS2, respectively. In this way, intra-spill signal ratios could be measured in favorable measurement conditions, where the maximum signal for each ion species is available.

Still, to analyze the temporal alignment between the beams, the helium signal needed to be isolated from the measured signal in IC3, which includes both helium ions and the fragment background of the carbon ions. We estimated the signal from the fragments from the difference of two IDD recorded with different helium concentrations. The signal in IC3 ($I_{IC3}$) is a linear combination of the carbon fragment background $F_C$ and the variable helium concentration $F_{He}$: $I_{(IC3,i)} = F_C + (x_i F_{He})$, where $x_i$, $i \in \{low, medium\}$ is the scaling factor according to the concentration. If $x_{med} = c\, x_{low}$, then

$$F_C = \frac{(cI_{(IC3,low)} - I_{(IC3,med)})}{(c-1)}$$

The factor $c$ was obtained by minimizing the residual of the IC3 measurements to an exponential fit $f(WET) = e^{(-a\, WET + b)}$ to the fragment tail of a pure carbon IDD obtained from Monte Carlo simulation, which was scaled to the measured carbon Bragg peak height and subtracted from IC3 measurements to derive the helium contribution. For any intra-spill measurement with IC2 in the carbon Bragg peak position, the function value was scaled by the instantaneous normalized IC2 signal.

The IDDs were replicated in Monte Carlo simulations conducted with the Geant4 toolkit (version 11.1). The setup consisted of a water phantom placed at isocenter, and the beam nozzle placed 1 m upstream. The beam lateral width at isocenter was adjusted to fit the beam profile measurements detailed in section B.2. The beam energy was set to the nominal beam energy during the experiments, with a 0.1% Gaussian energy spread. A 3D computer aided design (CAD) model of the 3 mm ripple filter used in Cave M was placed at the nozzle to modulate the beam energy spread, analog to the experiments [32]. Helium, carbon, and an oxygen contamination observed during the experiments were simulated separately. Per ion, $10^6$ primary particles were simulated. The ion's energy deposit to the water phantom was scored with a Geant4 primitive energy scorer. The "QGSP_BIC_EMZ" physics list was employed.

The verified simulations permit studying data not covered by the measurements. In particular, a calibration of the ICs positioned in the Bragg peak to the particle count was challenging, due to the sharp dose gradient and resulting high position sensitivity of the signal. This calibration factor was therefore obtained by fitting the Monte Carlo data to the measured IDDs.

*Contact author: C.Graeff@gsi.de

### 2. Beam profile measurements

Beam profile and beam divergence measurements were performed using Gafchromic EBT3 films (Ashland, Wilmington, DE, USA). Four films were used to measure the beam divergence of the mixed beam, placed at distances of 1.05 m, 1.69 m, 3.28 m, and 5.4 m to the nozzle. Two films were used to characterize the helium component, one at isocenter (1.05 m from the nozzle), and one at 1.69 m from the nozzle. For these, due to the dose difference between carbon and helium ions, four 50 mm PMMA blocks (232 mm WET) were positioned directly before the isocenter, such that the carbon ions were fully stopped inside the PMMA. The helium films were irradiated over several full spills to achieve sufficient signal.

The films were scanned with an EPSON Perfection V800 (EPSON, Japan) flatbed scanner at a resolution of 1200 dpi, and analyzed in the red color channel. Using the pixel values of an unirradiated film as reference, the scans were converted to optical density. The spot profile was then fitted with a double Gaussian to obtain the beam center and width of both beam components. In the films positioned behind the carbon ion Bragg peak, the broader beam component corresponds to the carbon ion fragment distribution. Due to the minimum dose needed for EBT3 films and the limited measurement time, the profile measurements were performed with the 'high' helium ion concentration setting. The beam divergence obtained for the carbon ion beam was used as input for a Monte Carlo simulation of the setup for both the carbon and helium beam component, which was then compared to the two helium film measurements to understand whether both ion types had a similar angular divergence.

### 3. Single-event measurements

To fully characterize the beam components, particle identification with a dE/E telescope was performed. The dE/E setup consisted of a 5 mm thick plastic scintillator (dE) placed in front of a BaF scintillator detector (E). The maximum count rate of this setup was 5 kHz, incompatible with clinical intensities used for the other tests. Hence, the beam intensity needed to be scaled to few kHz, which at GSI is performed through defocusing beyond the acceptance limit in the injector linac. This may change the concentration of helium ions with respect to that of carbon ions slightly, and therefore, cannot directly be compared against measurements at clinical intensity. In spite of a possible change in ratios at the reduced intensity, this setup enables the only direct separation of ion species.

*Contact author: C.Graeff@gsi.de

Due to the relationship between the energy loss in a thin dE and thick E absorber being a function of the mass and charge of the impinging ion species, plotting E versus dE enables to clearly separate different ion types. Ion species were identified through their signature in the dE/E plots using graphical cuts in Root (version 6.28) [24].

### C. Concurrent range verification

To demonstrate the potential of concurrent treatment and range monitoring in real time, we used two scintillator detectors operated in parallel: the first was an ion imaging scintillator prototype developed at University College London [25] (hereafter referred to as the UCL detector), the second one was developed at Centro Nazionale di Adroterapia Oncologica for mixed beam imaging [16], hereafter referred to as CNAO detector.

The UCL detector consisted of a 20 x 20 x 25 $cm^3$ Polystyrene scintillator (NUVIAtech, London, UK) read out by three high-speed cameras (BFS-U3-04S2M-CS USB3 cameras, Teledyne FLIR Integrated Imaging Solutions, Thousand Oaks, CA, USA). One camera captured the top view of the scintillator block, one the lateral, and one the distal view via a 45° angled mirror. This ion imaging detector has been extensively benchmarked at Marburg Ion Beam Therapy Centre for its capability of producing highly accurate ion radiographs [26], as well as for real-time tumor tracking [23]. The high-speed cameras could acquire one image per pencil beam, but a triggered acquisition introduced a deadtime of 2 ms after each image. This limitation of the prototype precluded the ability to directly record the full imaging field in sequence. Instead, every spot was irradiated twice, and an image was recorded only for every other spot.

The CNAO detector was a scintillator detector read out by a camera, too. In this case, a single, laterally mounted camera captured an image of the lateral view of the scintillation light output of the beam and the distal view via a 45° angled mirror simultaneously. The scintillator was a 20 x 20 x 20 $cm^3$ BC-408 plastic scintillator, and the Camera was an Andor Zyla 5.5 sCMOS. This camera took images at a fixed 200 ms framerate, with a deadtime of 0.1 ms in between, with the start triggered by the beam-on signal.

A ⌀6 cm PMMA sphere was placed at isocenter, and irradiated with an 80 x 28 $mm^2$ rectangular field with $5 \times 10^5$ carbon ions per spot at 2 mm spot spacing. The beam energy was 260 MeV/u, meaning the carbon ion beam stopped in the UCL detector, while the helium ions stopped in the CNAO detector. To provide a proof of concept, the helium concentration was kept as 'high' for this measurement, to avoid issues from current detector limitations. The acquired data of both systems was synchronized post hoc, comparing reconstructed beam lateral positions.

### D. Mixed-beam radiographic imaging

Using the UCL detector, spot-by-spot range verification was performed and full mixed beam radiographs were acquired for a phantom setup representing a lung tumor treatment. In detail, the setup consisted of 90 mm PMMA, i.e., 106.09 mm WET, representing the patient's proximal chest wall and soft tissue, that were positioned in front of a 30 mm diameter PMMA sphere (different to the 60 mm sphere used for the qualitative concurrent range monitoring before) on a linear stage (M404.2PD, Physik Instrumente, Karlsruhe, Germany), representing the moving tumor. Distal to the sphere, an additional 102.2 mm of PMMA, i.e., 117.5 mm WET, represented the patient's back. The mixed signal of the helium ions and carbon ion fragments behind the setup was recorded with the UCL detector. The detector entrance window was placed directly after the rear PMMA bolus, to minimize beam broadening. A photograph of the setup is provided in Figure 3.

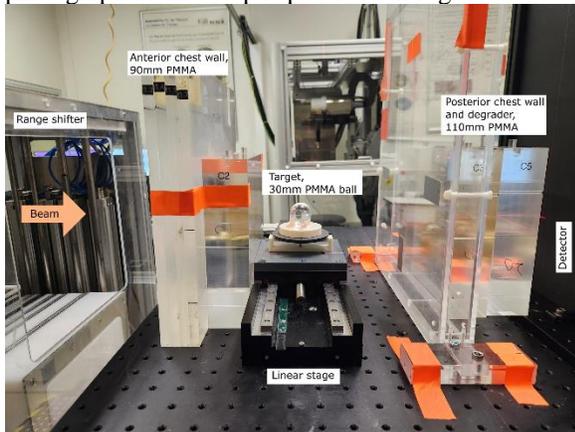

*Figure 3: Setup for mixed-beam portal imaging of a patient-like scenario. A PMMA ball placed at isocenter on top of a linear stage is used to simulate a lung tumor. PMMA blocks in front (90 mm) and behind (110 mm) the target acted as the chest wall, followed by the detector.*

The setup was irradiated with the same field as before but with $5 \times 10^6$ carbon ions per spot.

For image reconstruction, for every recorded pencil beam, the scintillation light output measured in the three cameras was corrected for optical artefacts, such as image distortion and reflections. Based on the corrected pencil beam, images were reconstructed using the ion imaging radiography framework developed by Simard et al. [27].

To analyze the potential of the mixed beam to provide the tumor position, images were acquired where the target was displaced by 1 mm, 2 mm, and 5 mm using the linear stage. The ball was segmented on the WET using Otsu's automatic threshold selection [28] and the contour's lateral center-of-mass position was compared to the set displacement. In addition, we simulated range changes, by inserting in sequence the first 3 plates of RS1 (0.5 mm, 1.24 mm, and 2 mm WET, respectively), and acquired an image for each. In addition to full images, the center-most spot of these images was used to quantify the ability of a mixed beam to provide spot-by-spot range verification in a complex setup.

### E. Assessing the clinical potential of mixed beams

Finally, the clinical potential of a mixed beam was investigated in silico, based on time-resolved CT (4DCT) data of 17 patients from the 4D-Lung dataset [29] available at the Cancer Imaging Archive [30].

A mixed beam treatment plan was generated for one of the patients using GSI's in-house treatment planning system TRiP98 [31-33]. The relative biological effectiveness (RBE) of the mixed beam was calculated based on the local effect model IV [34,35] assuming an alpha/beta ratio of 6Gy for the tumor and 2 Gy for healthy tissue. Two opposing (anterior-posterior, posterior-anterior) treatment fields delivering 4 Gy(RBE) were computed by 4D-robust optimization [36]. Both the carbon and helium components were considered, at a fixed percentage of 10% and 20% helium in the beam, using TRiP98's multi-ion RBE calculation [37,38]. The oxygen contamination currently present in the beam was not considered, as promising approaches for contaminant-free mixed beams are under investigation [20].

For all 17 patients, a Geant4 Monte Carlo simulation of the mixed-beam dose distribution in the patient and corresponding scintillation light output in a detailed model of the UCL detector described above was generated for ~19k pencil beams. Each pencil beam had a random direction, sampling from a gantry angle interval [30°,150°] and couch angle interval [30°,120°]. The pencil beam was then assigned a random position between -20 mm to 20 mm lateral displacement in horizontal and vertical direction in the beam's-eye-view (BEV) coordinate system. The beam energy was selected between 150 MeV/and 380 MeV/u, conditional that the helium range for the sampled parameters was sufficient to fully traverse the patient and still stop inside the range detector. The scintillation light quenching in the detector was analytically calculated using Birks' formula [39], at a quenching constant of 0.0089 cm/MeV. The simulation was performed on the reference phase of the 4DCT (end-exhale), as well as the end-inhale phase to study the correlation of carbon and helium range differences between those phases.

A deep learning model was developed on the Monte Carlo simulated pencil beam dataset, to produce a proof-of-concept for inferring carbon ion absolute range changes from the observed mixed beam detector signal. In fact, because helium integrates distal anatomy and density differences between detector and patient, and range changes are observed in a detector that differs in density compared to the various tissues of the patient anatomy crossed by the carbon ion beam, a direct physics-based mapping is insufficient, motivating an artificial intelligence approach. A convolutional network is the most suited for this task, as it can directly handle the output of our current imaging setup with a scintillator observed by multiple cameras.As fixed input, the model received a BEV projection of the reference phase of the 4DCT (end-exhale), the 3D dose distribution of the mixed beam on the BEV CT, the three 2D projections (top, lateral, and distal view) of the scintillation light output of the detector calculated on reference phase, as well as the nominal energy and width of the beam. As the variable input, the model received the three projections of the scintillator light output as would be observed in real-time during treatment, which here corresponded to the simulation for the end-exhale phase of the 4DCT, i.e., the furthest position shift compared to reference. The helium scintillation light output was preprocessed to remove the carbon ion fragment background and noise through a U-Net based image2image model trained on paired simulation data of pure helium and mixed beam scintillator images (further details provided as supplementary material).

*Contact author: C.Graeff@gsi.de

The BEV CT and reference dose distribution were processed through a 3D convolutional encoder, the two sets of scintillator images through a 2D convolutional encoder, and the two scalar values through a multi-layer-perceptron. The output of these was concatenated, resulting in a latent feature vector of size 512. This was then processed through a multi-layer perceptron to output the carbon ion's 80% dose fall off (R80) shift. The output of the range inference model was the shift of the carbon ion beam Bragg peak position inside the patient, expressed as the radiological distance between the R80 between end-inhale and end-exhale.

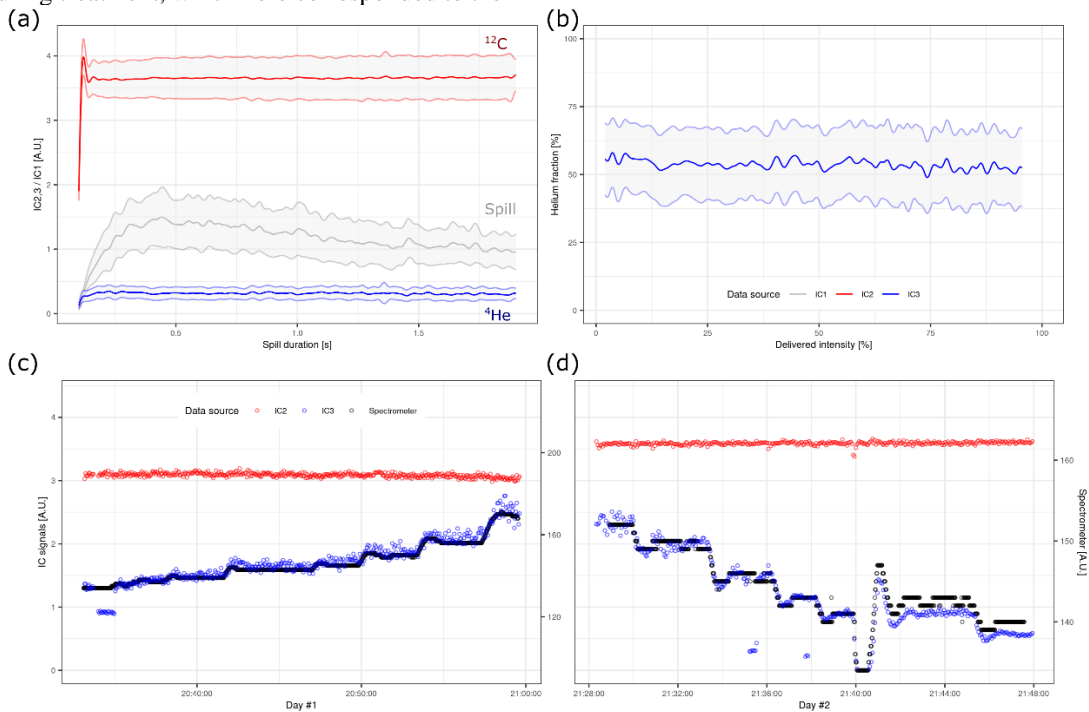

*Figure 4:* (a)-(b): Time-resolved data of IC1-3 during the spill. (a) shows the average over 21 consecutive spills, with ±1 SD shown as the shaded area. IC2 and IC3 are normalized to IC1, IC3 was corrected for the higher amplification of a factor of 50 compared to IC2. (b) shows the estimated helium concentration derived from the same data, where the carbon fragment background has been subtracted. (c)-(d): Correlation of the deliberate variation of the helium injection in the source, measured as analyzed beam current in the low-energy beam transport, with extracted helium concentration in Cave M. The helium Bragg peak is measured in IC3 after correction for carbon fragment background and normalization to IC1. The carbon Bragg peak signal in IC2 varies only minimally with the strongly changing helium concentration in the beam. Each data point represents the integral signal over one spill.

*Contact author: C.Graeff@gsi.de

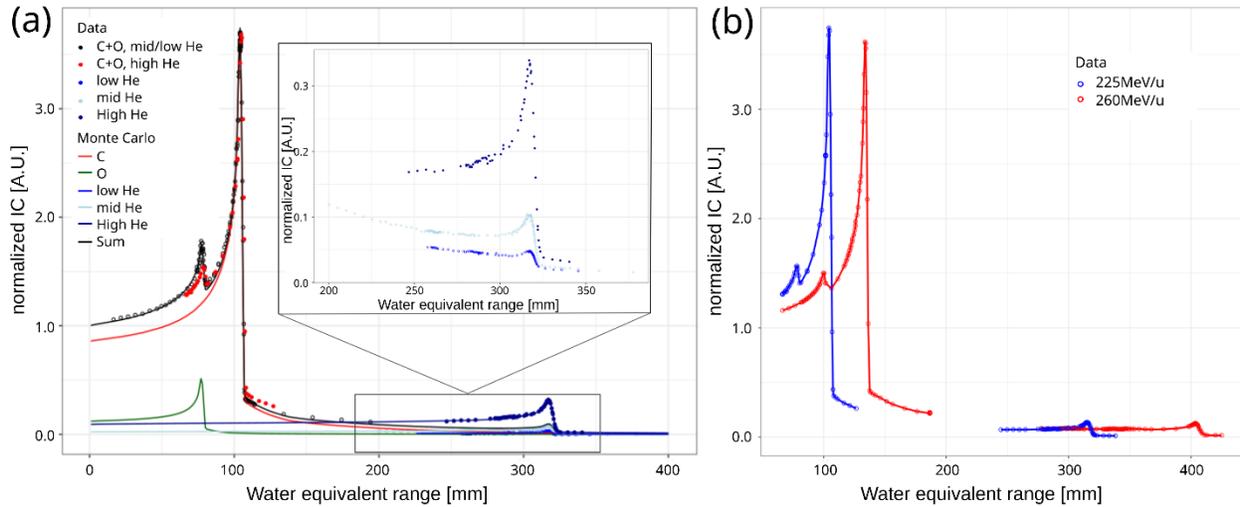

*Figure 5:* (a) Full depth-dose curve of the mixed beam at 225 MeV/u for different helium concentrations highlighting the clean separation of the helium Bragg peak from the carbon fragment tail even at low helium concentrations. The oxygen and carbon data points were acquired with IC2, while the datapoints corresponding to the helium ions and the fragment tail were measured with IC3. Due to the weaker signal of the helium compared to the primary carbon ions, compared to IC1 and IC2, the signal of IC3 was amplified a factor 50 more for the 'low' and 'medium', and a factor 20 for 'high' helium setting. The datapoints were adjusted to the same normalization. The inset plot shows a detail view on the datapoints measured with IC3. (b) Two IDD acquired as described for (a), showing two different incident beam energies, 225 MeV/u and 260 MeV/u. Notably, the factor 3 between the carbon and helium range difference can be seen.

## III. RESULTS

### A. Beam characterization
#### 1. Mixed beam stability

To study the temporal behavior of the carbon and helium concentration, IC2 and IC3 were positioned in the Bragg peak of C and He, respectively. IC3 was corrected for carbon fragments as described above. In this setting, both interspill and intraspill variability was studied, for interspill variability, only integrated values per spill are reported. Figure 4 (a) and (b) show the intraspill ratios of helium and carbon measured for the 'high' helium concentration at 50 kHz sampling frequency over the spill duration of 2 s. In the figure, the nozzle signal is included to show the general spill shape at GSI. The spill profile was relatively flat. Remarkably, a stable ratio between helium and carbon ions was achieved over the full spill. The stability of the helium ion concentration is ideal for concurrent treatment and range monitoring, where the expected helium signal in the range measuring detector can then be determined pre-treatment.

On two successive days, the correlation of manual adjustment of the injection valve in the ECRIS to the helium signal in Cave M was monitored, as depicted in Figure 4(c) and (d). The reference value is the analyzed beam current measured in the low-energy beam transport, acting as a mass spectrometer, behind the ion source. At the given energy per nucleon of 2.5 keV/u, all ions with A/Q=4 are measured; more contaminants could possibly be present in the ion beam (momentum to charge selection). This data was matched to the IC data from Cave M. As the two DAQ systems were not running on a synchronized clock, the time axis was shifted in both runs by 25 s to align the measurements. The relationship of beam current to IC data varies between both runs, which differed in extraction mode. The first run was taken prior to tuning the extraction, such that carbon and helium were extracted subsequently [21]. Nevertheless, the helium signal in Cave M follows the source signal closely over a large variation of helium concentration. It can also be seen that the carbon signal in IC2 is only marginally affected by approximately doubling the helium content. This illustrates that the IC1 signal, to which this data is normalized, is dominated by carbon. This is a significant finding for clinical application, as the delivered treatment dose is therefore robust against fluctuations in the helium concentration.

*Contact author: C.Graeff@gsi.de

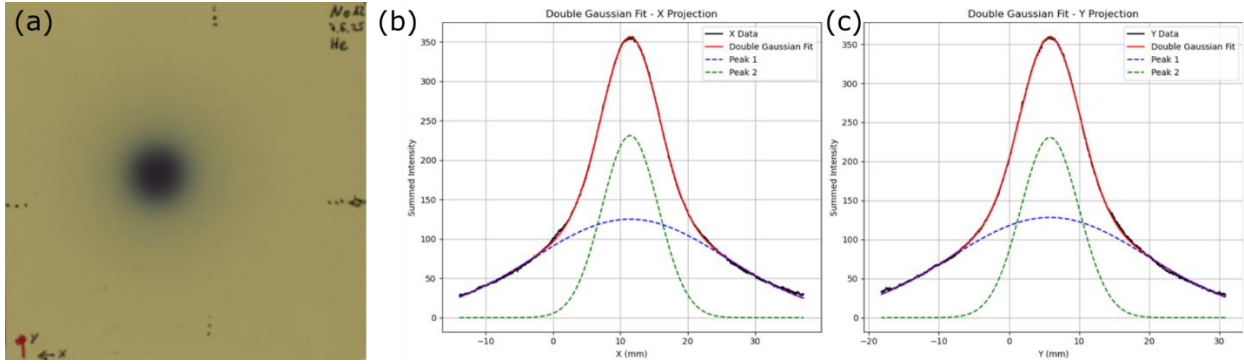

*Figure 6: (a) scan of a Gafchromic EBT3 film placed behind 200 mm PMMA and irradiated by the mixed beam. The dose stems from the helium ions, as well as from the fragments of the primary carbon ion beam. (b) Integral horizontal profile of the film optical density. (c) Integral vertical profile. A double-Gaussian fit (red) to the profile, and its two individual components, representing the helium ions (green, dashed) and carbon fragments (blue, dashed), are shown. For (b), the fit parameters were $\mu_{He}$ = 11.45 ± 0.01 mm, $\sigma_{He}$ = 4.18 ± 0.01 mm, $\mu_C$ = 11.42 ± 0.01 mm, $\sigma_C$ = 14.29 ± 0.02 mm. For (c) they were $\mu_{He}$ = 5.84 ± 0.01 mm, $\sigma_{He}$ = 4.15 ± 0.01 mm, $\mu_C$ = 5.83 ± 0.01 mm, $\sigma_{He}$ = 14.10 ± 0.02 mm.*

### 2. Depth-dose distribution

Figure 5 shows the IDD as recorded by IC2 and IC3 for the three different helium concentrations. Each datapoint is the average value of several spills (minimum 5, median 7), acquired in several runs of 2 - 3 h each. From these measurements, the carbon fragment background was derived, using 174 data points in the overlapping region of the IC3 curves. The coefficient found was c=3.11, resulting in $f(WET) = e^{(-0.0127\,WET+1.5099)}$, with RMSE = 0.005 and R² = 0.98. In the right panel of Figure 2, helium IDDs are shown as a result of subtracting the fitted carbon fragment background. Moreover, the IDD obtained through Geant4 MC simulation is overlayed to the data, showing the beam components C, He, and O individually and as a sum of all contributions. To fit IC3, the MC data was scaled by the difference in amplification. All MC data was scaled to the maximum of IC2, and relative concentrations of O and He were selected to match the data. The concentrations for the best fit are 100% C, 7.9% O and 7.7% or 24.0% He for the low and medium helium concentration runs. For the 'high' helium concentration, it was 100%C, and 93% He, and 4.9% O, i.e., ~47% of the beam was helium.

### 3. Beam profile and envelope

Using the four Gafchromic EBT3 films placed at different distances to the nozzle, the primary beam FWHM was determined to be 5.99 mm in horizontal and 5.11 mm in vertical direction at isocentre, with an angular divergence of $\sigma_\theta$ = 2.6 mrad (determined through analytical modeling of the scattering [40,41], see supplementary material). These measurements are dominated by the carbon ion beam and well comparable to the typical beam width and divergence at carbon ion therapy facilities [42]. Figure 6 shows a Gafchromic EBT3 film positioned behind 200 mm of PMMA, i.e., distal to the carbon ion Bragg peak, such that the helium beam is visible over the carbon fragment background. The double Gaussian fit to the data, and the two individual Gaussians representing the carbon and helium beam are shown. The distance of the centers of the fitted Gaussians for He and C was dx = (0.03 ± 0.01) mm and dy = (0.01 ± 0.01) mm, where the given uncertainty is that of the fit. The isocentric beam at GSI Cave M requires a deflection of x=100 mm and y=300 mm from the beam position at the scanning magnets. For perfectly aligned beams at the scanner magnets, the expected shifts due to the mass defect would be 0.065 mm and 0.20 mm for x and y, respectively. Although the error is large compared to the uncertainty of the Gaussian fit to the data, considering the inherent uncertainty of dose measurements with radiosensitive films [43], we conclude that both beams were well aligned.

The FWHM of the fits were (9.8 ± 0.01) mm for helium and (33.7 ± 0.01) mm for the carbon fragments. Replicating the setup for helium film measurements in a Monte Carlo simulation and assuming the beam divergence of the carbon ion beam also for the helium ions matched the observed helium films well. We therefore conclude that both beams had a similar beam divergence of $\sim\sigma_\theta$ = 2.6 mrad before the nozzle beam monitors.

*Contact author: C.Graeff@gsi.de

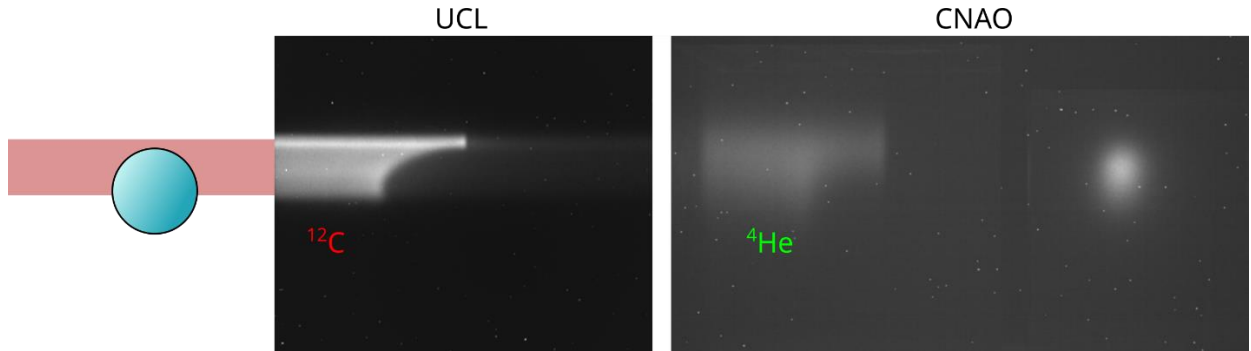

*Figure 7:* Example frame of simultaneous data acquisition with the UCL and CNAO scintillator detectors. The mixed beam was scanned over a 6 cm PMMA ball, after which the carbon ions stopped inside the UCL detector, while the helium ions stopped in the CNAO detector. The shown frame integrates several pencil beams over a 200 ms interval, due to the sampling frequency of the CNAO scintillator. The UCL detector shows the lateral view, the CNAO detector simultaneously the lateral and distal view of the scintillation light produced by the beam. The full video of the acquisition can be found in the Supplementary Material.

*4. Single-event detection*

The variation of the helium concentration over the spill as measured with the dE/E telescope agreed well with that measured with the IC stack setup. The concentration of oxygen was stable over the spill. As the mass/charge ratio between fully stripped oxygen and carbon ions differs even less than that between helium and carbon ions, the stability of oxygen with respect to carbon is somewhat expected. The average ratios for helium and oxygen with respect to carbon were 18.8% and 6.2%, respectively, in the 'medium' helium setting, and 4.7% and 7.3%, respectively, in the 'low' helium setting. The difference to the IC measurement contamination estimate likely is a result of the difference in beam current (O(10 MHz) versus O(1 kHz)), as well as difference in the measurement setup.

**B. Concurrent range verification**

A short video of the aligned simultaneous measurements of the carbon ions stopping inside the UCL detector and the helium ions stopping inside the CNAO detector has been generated, and can be found in the supplementary material of this manuscript. In Figure 7, we provide two example frames from this video, to highlight the potential of the simultaneous treatment and range monitoring. The helium ion ranges observed in the CNAO detector qualitatively match the carbon ion signal in the UCL detector very well: the shape of the PMMA ball is visible both in the carbon ion and the helium range pull-back. Particularly, the top-most pencil beam that passes the sphere unobstructed is clearly visible in both detectors. A quantitative analysis of spot-by-spot range monitoring is provided in section III.C.

**C. Mixed-beam radiography and spot-by-spot range probing**

Figure 8(a) shows the first mixed beam radiographs acquired with the scintillator detector with the mixed beam in the 'high' helium ion setting. The top image in Figure 8(a) shows the reconstructed image of the lung-cancer mimicking setup with the PMMA tumor in the isocenter position. The 30 mm diameter ball is clearly visible in the image, and its outline could be reconstructed automatically with a grey value thresholding algorithm. The reconstructed tumor center of mass is indicated as a small cross on the image. The other images in Figure 8(a) show the reconstructed mixed beam radiographs for three tumor position shifts. The true and measured horizontal shift of the tumor centroid, as well as the difference in WET in a central region of the ball of 1 cm diameter reconstructed from the mixed beam radiographs are provided in Table 1. Values generally agreed very well with the ground truth, with tumor position accuracy well below the pixel size with a mean accuracy of 0.19 mm and a range shift accuracy below 0.1% of the total WET (256 mm) of the setup. Figure 8(b) and Figure 8(c) show integrated scintillation light profiles of the lateral view of the UCL detector signal for the center-most pencil beam from the imaging field, corrected for the carbon ion fragment background by the developed U-Net model. In both graphs, the red curve corresponds to the reference acquisition, while the blue curve is from an acquisition with an additional 1.24 mm WET plate and 0.49 mm WET plate in front of the setup, respectively. These shifts mimic realistic range shift scenarios for a clinical patient. The observed R80 shift was 1.38 mm and 0.44 mm, demonstrating the capability of spot-by-spot range verification with the mixed beam setup.

*Contact author: C.Graeff@gsi.de

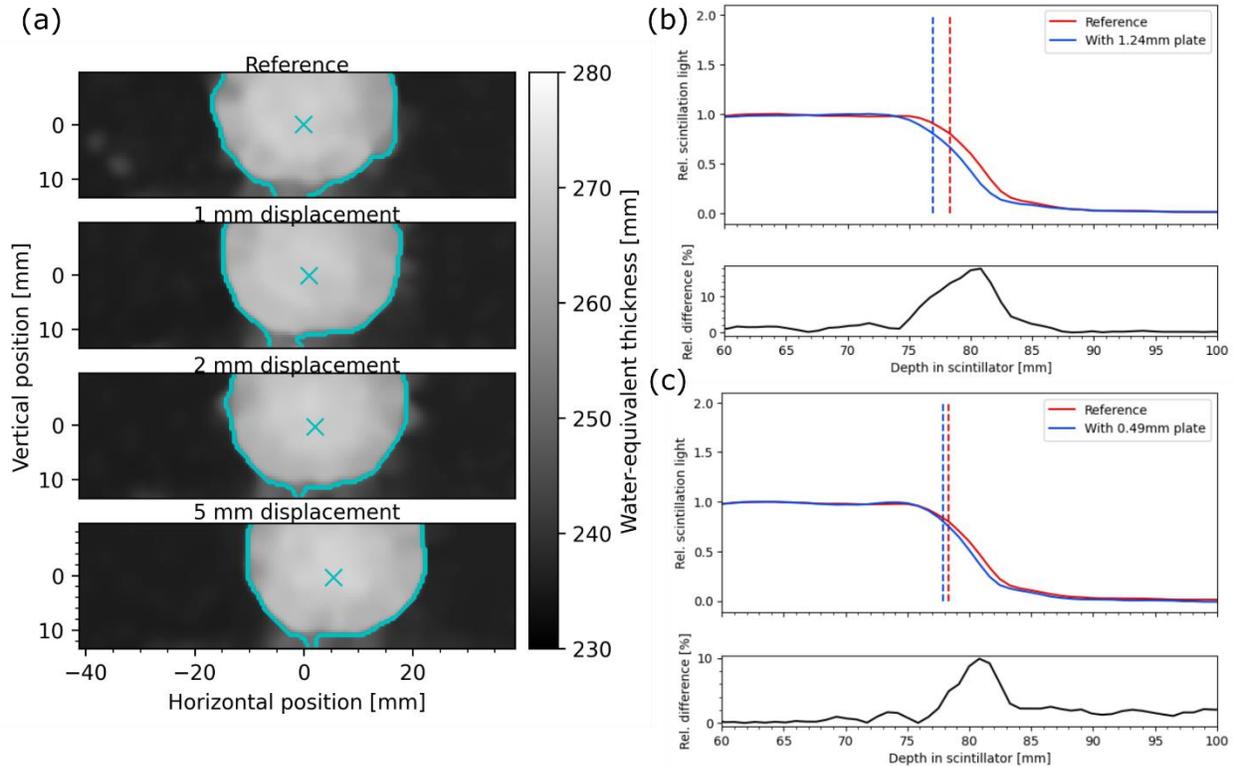

***Figure 8:*** *(a) Mixed beam radiographs of a 3 cm diameter tumor in a lung-cancer-like setup. The top image shows the reference image, while the other four images show the tumor laterally displaced with a linear stage. The cyan line shows the automatic tumor segmentation, and the cross the reconstructed centroid. (b) and (c) show the integrated lateral view of the UCL detector scintillation light output for the center-most pencil beam of the imaging field. The red Bragg curve shows the pencil beam for the reference acquisition. The blue curve in (b) is from an acquisition with a 1.24 mm WET plate in front of the setup, while the blue curve in (c) is from an acquisition with a 0.49 mm WET plate in front of the setup. The dashed lines indicate the reconstructed R80. The bottom part of the graph shows the relative difference between the shifted and reference Bragg curve. The mixed beam signal in the UCL detector was corrected for the carbon fragment background, and single-pixel noise was removed.*

***Table 1:*** *Position and range shifts of the tumor reconstructed from the mixed beam radiographs acquired with the UCL detector in comparison to the ground truth.*

| Tumor shift [mm] | Reconstructed centroid shift [mm] | Absolute error [mm] |
|---|---|---|
| 1.0 | 0.98 | 0.02 |
| 2.0 | 2.08 | 0.08 |
| 5.0 | 5.47 | 0.47 |
| Added WET [mm] | Reconstructed WET shift [mm] | Absolute error / relative to total WET |
| 0.49 | 0.38 | 0.11 mm / 0.05% |
| 1.24 | 1.46 | 0.22 mm / 0.09% |
| 2.0 | 2.12 | 0.12 mm / 0.05% |

### D. In-silico assessment of clinical potential

Figure 9(a) shows the treatment plan generated for a lung cancer patient with the mixed beam at a 10% helium concentration. Figure 9(b) shows the contribution of only the helium ion beam to the biological dose, computed by subtracting forward dose calculations with and without the helium contamination from each other. At 10% concentration, the helium adds about 1% to the physical dose in the target, which corresponds to less than 0.5% of the RBE-weighted dose. At 20%, the helium ion contribution is at about 2% of the physical and less than 1% of the RBE-weighted dose.

Figure 9(c) shows the correlation of the difference in helium range inside the detector between the end-inhale and end-exhale (reference) phase and the corresponding difference in carbon range inside the patient obtained with Geant4 Monte Carlo simulations.

*Contact author: C.Graeff@gsi.de

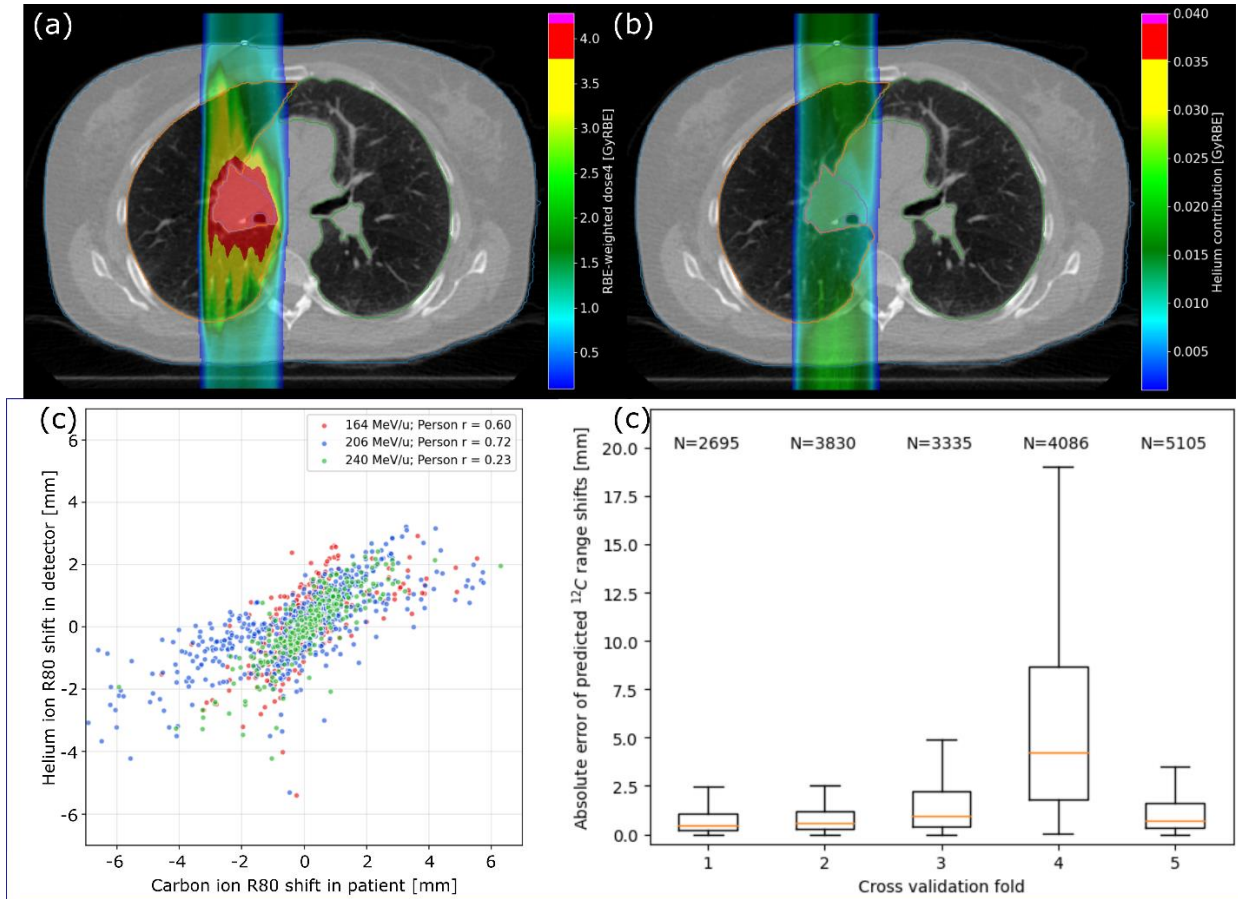

*Figure 9:* *(a) Mixed beam treatment plan for a patient with a centrally located lung tumor. (b) Contribution of the biological dose stemming from just the helium contamination. Inside the target, it was <0.5% of the target biological dose. (c) Correlation of the carbon ion and helium ion range difference between end-inhale and end-exhale (reference) phase of the 4DCT for a proximal, central, and distal energy layer from the anterior-posterior field from the plan shown in (a). The data was obtained by a Geant4 simulation of the patient and detector setup. Pearson's r was computed as a measure of correlation between the helium and carbon Bragg peak position shifts. (d) Distribution of the error of the deep-learning predicted difference in the carbon ion range from the scored scintillation light output of the helium beam inside the detector across 17 patients (14 used for training and 3 for validation per fold), at 19k simulated pencil beams. N indicates the number of pencil beams for the three validation patients.*

In Table 2, we report the overall performance of our deep learning model in predicting the range difference for the random sampled beam spots across the 5-fold cross validation (N=19051 pencil beams). Spot-by-spot range prediction had a mean accuracy of 2.9 mm across the five folds, with fold 4 yielding the lowest accuracy. Overall, the model prediction showed an on-average improvement of 44% compared to assuming no range changes in the carbon ion beam, and an on average 45% improvement compared to a naïve direct approach, assuming range differences in helium directly correlate to carbon ion range differences. Detailed plots on the model performance are provided as supplementary material. Figure 9(d) shows the distribution of the spot-by-spot predicted carbon ion range difference from the observed helium range difference for the specific case of the treatment plan shown in Figure 9(a).

*Table 2:* *Evaluation of the spot-by-spot accuracy of the carbon ion range difference between the end-inhale and end-exhale breathing phase predicted from the helium ion scintillator signal from 19k Geant4 Monte Carlo simulated mixed beam samples across 17 patients. The 5 cross validation folds contain the data from 14 of the patients for training, while the data from the three remaining patients was reserved for validation.*

| Fold | MAE of predicted carbon R80 [mm] | Improvement to assuming no range changes | Improvement to translating helium range changes directly to carbon |
|---|---|---|---|
| 1 | 1.62 | 40% | 69% |
| 2 | 1.27 | 46% | 12% |
| 3 | 2.52 | 59% | 74% |
| 4 | 6.60 | 33% | 10% |
| 5 | 2.54 | 41% | 61% |

*Contact author: C.Graeff@gsi.de

## IV. DISCUSSION

The results demonstrate, for the first time, the controlled production and use of mixed carbon–helium ion beams for image-guided heavy-ion therapy. The experiments establish three key results: (i) stable simultaneous acceleration and extraction of carbon and helium ions at clinically relevant energies, (ii) reliable separation of the helium signal from the carbon background for range probing, and (iii) proof-of-concept portal imaging of a tumor-like phantom with sub-millimeter accuracy. Together, these findings show that mixed beams can simultaneously deliver therapeutic dose and provide real-time anatomical and range information—a capability long considered a "missing link" in ion therapy.

The clinical implications are significant. Current heavy-ion treatments suffer from range uncertainty, forcing the use of conservative margins that undermine one of the central advantages of carbon therapy. Techniques based on secondary radiation or anatomical imaging remain constrained either by low signal-to-noise or by the indirect inference of range. By contrast, the mixed beam approach provides direct, actionable range information with negligible additional dose, while simultaneously enabling portal imaging, a standard in photon therapy but so far absent in ions. In a clinical workflow, this information could be used for online beam interruption, adaptive re-scanning, or margin reduction without additional imaging dose. In this respect, mixed beams offer not just incremental improvement but a paradigm shift in how image guidance can be integrated into carbon ion therapy. For example, helium ions added negligible dose to the treatment plan shown in Figure 9, while the added information of the helium beam enables highly accurate range inference, as demonstrated in silico with deep-learning inference (Table 2), and experiment (Figure 7). Of note is that the helium ion range being conditional on the carbon ion range favors certain beam directions over others: For the patient investigated in Figure 9 and field configuration, most of the helium ions had enough range to fully traverse the patient for all spots. However, the same would have not been the case, if a lateral field was chosen. This issue can easily be overcome by adding a bolus in front of the patient, as recently proposed in the simulation study by Hardt et al. [44].

Nevertheless, the correlation between helium and carbon ion suffers from the density difference between the scintillator and patient tissue, as well as the fact that the helium beam integrates also all range changes distal to the target. Although plastic scintillator is an excellent detector for quality assurance in particle therapy owing to its near-equivalence to soft tissue [45], predicting the comparatively large range shifts when the carbon ion beam crosses into lung tissue presents a challenge. Lung tissue has a factor of ~3 lower density compared to soft tissue. If the solid tumor moves out of the beam path due to, e.g., breathing motion, the corresponding physical change in the Bragg peak position of the carbon beam would be substantial, while it would be about three times less for the helium beam inside the detector. In contrast, the helium beam range inside the detector would also shift for any anatomical change occurring distal to the target, even if the carbon ion Bragg peak position remains unchanged. This scenario has to be adequately addressed as well, to prevent false positives.

We have shown that a highly accurate range inference can be accomplished with a deep learning model that gets a BEV CT, the reference carbon dose and reference scintillator images, as well as the recorded scintillator signal. Of note is that simpler models, such as an MLP acting only on the helium range difference did yield reduced performance, and prior information on the patient geometry such as the BEV CT used here appears necessary. In addition, while the data presented here can be understood as a solid proof of concept, patient individual differences have to be studied in more detail. For example, the result of one of the cross-validation folds of our model was negatively influenced by a single patient from the dataset where larger carbon ion range differences with respect to the overall distribution across all patients were observed.

The present demonstration also exposes important technical challenges. Leading among these is the oxygen contamination observed in the beam, which must be eliminated for clinical translation. This issue is tractable: alternative ion-source strategies or multi-turn injection schemes can yield high-purity beams. For example, recent work at the Austrian carbon ion therapy center MedAustron (Wiener Neustadt, Austria), Kausel et al. [20] have successfully generated a mixed beam with a double-multi-turn injection scheme. Helium and carbon ions are injected into the accelerator from different sources one after another, which eliminates any possible beam contamination.

Importantly, the core concept of mixed-beam feasibility and its use for treatment monitoring is firmly established through this work. Beyond oncology, the generation of mixed ion beams at synchrotrons opens opportunities across physics. For plasma physics, one ion species could serve as a driver while another probes plasma evolution. In nuclear physics, mixed beams may provide in-situ calibration or precision energy measurements. These perspectives underscore that the significance of mixed beams extends beyond medicine, positioning them as a novel tool for a range of accelerator-based sciences.

*Contact author: C.Graeff@gsi.de

## V. CONCLUSIONS

The first realization of real-time range monitoring and imaging with a mixed carbon–helium beam represents a decisive step toward image-guided heavy-ion therapy. By uniting treatment and real-time verification within the same beam, this approach directly addresses one of the field's most critical limitations and paves the way toward exploiting the full therapeutic potential of carbon ions. Future work will focus on optimizing beam production, developing dedicated detector systems, and translating the method into pre-clinical and eventually clinical studies.

## ACKNOWLEDGMENTS

Funded by the European Union (ERC, PROMISE, 101124273). This project has received funding from the European Union's Horizon 2020 research and innovation programme under grant agreement No 101124273. Views and opinions expressed are however those of the author(s) only and do not necessarily reflect those of the European Union or the European Research Council Executive Agency. Neither the European Union nor the granting authority can be held responsible for them.

The experiments were excellently supported by R. Khan and C. Hartmann-Sauter, both GSI.

## APPENDIX

A supplementary material is supplied with this manuscript, containing supporting information on the setup, methods, and a multi-media results file.

## REFERENCES

[1] C. Graeff, L. Volz, and M. Durante, *Emerging technologies for cancer therapy using accelerated particles*, Prog Part Nucl Phys **131**, 104046 (2023).

[2] M. Durante, R. Orecchia, and J. S. Loeffler, *Charged-particle therapy in cancer: clinical uses and future perspectives*, Nature Reviews Clinical Oncology **14**, 483 (2017).

[3] S. van de Water, S. Safai, J. M. Schippers, D. C. Weber, and A. J. Lomax, *Towards FLASH proton therapy: the impact of treatment planning and machine characteristics on achievable dose rates*, Acta Oncologica **58**, 1463 (2019).

[4] J. Y. Chang et al., *Consensus Guidelines for Implementing Pencil-Beam Scanning Proton Therapy for Thoracic Malignancies on Behalf of the PTCOG Thoracic and Lymphoma Subcommittee*, Int J Radiat Oncol Biol Phys **99**, 41 (2017).

[5] A. M. Chhabra et al., *Proton Beam Therapy for Pancreatic Tumors: A Consensus Statement from the Particle Therapy Cooperative Group Gastrointestinal Subcommittee*, Int J Radiat Oncol Biol Phys **122**, 19 (2025).

[6] T. Pfeiler et al., *Motion effects in proton treatments of hepatocellular carcinoma-4D robustly optimised pencil beam scanning plans versus double scattering plans*, Phys Med Biol **63**, 235006 (2018).

[7] A. C. Knopf and A. Lomax, *In vivo proton range verification: a review*, Phys Med Biol **58**, R131 (2013).

[8] K. Parodi and J. C. Polf, *In vivo range verification in particle therapy*, Med Phys **45**, e1036 (2018).

[9] K. Parodi, *Vision 20/20: Positron emission tomography in radiation therapy planning, delivery, and monitoring*, Med Phys **42**, 7153 (2015).

[10] A. C. Kraan et al., *In-beam PET treatment monitoring of carbon therapy patients: Results of a clinical trial at CNAO*, Phys Med **125**, 104493 (2024).

[11] J. Berthold et al., *First-In-Human Validation of CT-Based Proton Range Prediction Using Prompt Gamma Imaging in Prostate Cancer Treatments*, Int J Radiat Oncol Biol Phys **111**, 1033 (2021).

[12] F. Hueso-Gonzalez, M. Rabe, T. A. Ruggieri, T. Bortfeld, and J. M. Verburg, *A full-scale clinical prototype for proton range verification using prompt gamma-ray spectroscopy*, Phys Med Biol **63**, 185019 (2018).

[13] D. Boscolo et al., *Image-guided treatment of mouse tumours with radioactive ion beams*, Nature Physics (2025).

[14] S. Mori et al., *Commissioning of a fluoroscopic-based real-time markerless tumor tracking system in a superconducting rotating gantry for carbon-ion pencil beam*

*Contact author: C.Graeff@gsi.de

scanning treatment, Med Phys **46**, 1561 (2019).

[15] S. Gantz, L. Karsch, J. Pawelke, J. Schieferecke, S. Schellhammer, J. Smeets, E. van der Kraaij, and A. Hoffmann, *Direct visualization of proton beam irradiation effects in liquids by MRI*, Proc Natl Acad Sci U S A **120**, e2301160120 (2023).

[16] D. Mazzucconi, S. Agosteo, M. Ferrarini, L. Fontana, V. Lante, M. Pullia, and S. Savazzi, *Mixed particle beam for simultaneous treatment and online range verification in carbon ion therapy: Proof-of-concept study*, Med Phys **45**, 5234 (2018).

[17] L. Volz et al., *Experimental exploration of a mixed helium/carbon beam for online treatment monitoring in carbon ion beam therapy*, Phys. Med. Biol. **65**, 055002 (2020).

[18] J. J. Hardt, A. A. Pryanichnikov, N. Homolka, E. A. DeJongh, D. F. DeJongh, R. Cristoforetti, O. Jakel, J. Seco, and N. Wahl, *The potential of mixed carbon-helium beams for online treatment verification: a simulation and treatment planning study*, Phys Med Biol **69** (2024).

[19] M. Galonska et al., *First dual isotope beam production for simultaneous heavy ion radiotherapy and radiography*, 1893 (2024).

[20] M. Kausel et al., *Double multiturn injection scheme for generating mixed helium and carbon ion beams at medical synchrotron facilities*, Physical Review Accelerators and Beams **28** (2025).

[21] D. Ondreka, L. Bozyk, C. Graeff, P. Spiller, J. Stadlmann, and L. Volz, *Slow extraction of a dual-isotope beam from SIS18*, 1698 (2024).

[22] F. Maimone, J. Mader, R. Lang, P. T. Patchakui, K. Tinschert, and R. Hollinger, *Optical spectroscopy as a diagnostic tool for metal ion beam production with an ECRIS*, Rev Sci Instrum **90**, 123108 (2019).

[23] R. Fullarton et al., *Imaging lung tumor motion using integrated-mode proton radiography-A phantom study towards tumor tracking in proton radiotherapy*, Med Phys **52**, 1146 (2025).

[24] R. Brun and F. Rademakers, *ROOT — An object oriented data analysis framework*, Nuclear Instruments and Methods in Physics Research Section A: Accelerators, Spectrometers, Detectors and Associated Equipment **389**, 81 (1997).

[25] M. Lis, W. Newhauser, M. Donetti, M. Durante, U. Weber, B. Zipfel, C. Hartmann-Sauter, M. Wolf, and C. Graeff, *A facility for the research, development, and translation of advanced technologies for ion-beam therapies*, Journal of Instrumentation **16** (2021).

[26] M. Simard et al., *A comparison of carbon ions versus protons for integrated mode ion imaging*, Med Phys **52**, 3097 (2025).

[27] M. Simard, D. G. Robertson, R. Fullarton, G. Royle, S. Beddar, and C. A. Collins-Fekete, *Integrated-mode proton radiography with 2D lateral projections*, Phys Med Biol **69** (2024).

[28] N. Otsu, *A Threshold Selection Method from Gray-Level Histograms*, IEEE Transactions on Systems, Man, and Cybernetics **9**, 62 (1979).

[29] G. D. Hugo, E. Weiss, W. C. Sleeman, S. Balik, P. J. Keall, J. Lu, and J. F. Williamson, *A longitudinal four-dimensional computed tomography and cone beam computed tomography dataset for image-guided radiation therapy research in lung cancer*, Med Phys **44**, 762 (2017).

[30] The Cancer Imaging Archive, www.cancerimagingarchive.net.

[31] M. Krämer, O. Jäkel, T. Haberer, G. Kraft, D. Schardt, and U. Weber, *Treatment planning for heavy-ion radiotherapy: physical beam model and dose optimization*, Phys. Med. Biol. **45**, 3299 (2000).

[32] S. Hild, C. Graeff, A. Rucinski, K. Zink, G. Habl, M. Durante, K. Herfarth, and C. Bert, *Scanned ion beam therapy for prostate carcinoma : Comparison of single plan treatment and daily plan-adapted treatment*, Strahlenther Onkol **192**, 118 (2016).

[33] K. Anderle, J. Stroom, S. Vieira, N. Pimentel, C. Greco, M. Durante, and C. Graeff, *Treatment planning with intensity modulated particle therapy for multiple targets in stage IV non-small cell lung cancer*, Phys Med Biol **63**, 025034 (2018).

*Contact author: C.Graeff@gsi.de

[34] T. Friedrich, U. Scholz, T. Elsässer, M. Durante, and M. Scholz, *Calculation of the biological effects of ion beams based on the microscopic spatial damage distribution pattern*, International Journal of Radiation Biology **88**, 103 (2011).

[35] T. Pfuhl, T. Friedrich, and M. Scholz, *Comprehensive comparison of local effect model IV predictions with the particle irradiation data ensemble*, Med Phys **49**, 714 (2022).

[36] T. Steinsberger et al., *Evaluation of motion mitigation strategies for carbon ion therapy of abdominal tumors based on non-periodic imaging data*, Phys Med Biol **70** (2025).

[37] E. Scifoni, W. Tinganelli, W. K. Weyrather, M. Durante, A. Maier, and M. Kramer, *Including oxygen enhancement ratio in ion beam treatment planning: model implementation and experimental verification*, Phys Med Biol **58**, 3871 (2013).

[38] W. Tinganelli, M. Durante, R. Hirayama, M. Kramer, A. Maier, W. Kraft-Weyrather, Y. Furusawa, T. Friedrich, and E. Scifoni, *Kill-painting of hypoxic tumours in charged particle therapy*, Sci Rep **5**, 17016 (2015).

[39] J. B. Birks, *Scintillations from Organic Crystals: Specific Fluorescence and Relative Response to Different Radiations*, Proceedings of the Physical Society. Section A **64**, 874 (1951).

[40] B. Gottschalk, *Techniques of Proton Radiotherapy: Transport Theory*, **arXiv:1204.4470** (2012).

[41] B. Gottschalk, *On the scattering power of radiotherapy protons*, Med Phys **37**, 352 (2010).

[42] C. Kleffner, D. Ondreka, and U. Weinrich, *The Heidelberg Ion Therapy (HIT) Accelerator Coming into Operation*, Aip Conf Proc **1099**, 426 (2009).

[43] R. Castriconi et al., *Dose–response of EBT3 radiochromic films to proton and carbon ion clinical beams*, Physics in Medicine and Biology **62**, 377 (2016).

[44] J. J. Hardt, A. A. Pryanichnikov, O. Jakel, J. Seco, and N. Wahl, *Helium range viability for online range probing in mixed carbon-helium beams*, Med Phys **52**, e70194 (2025).

[45] A. S. Beddar, K. Kainz, T. M. Briere, Y. Tsunashima, T. Pan, K. Prado, R. Mohan, M. Gillin, and S. Krishnan, *Correlation between internal fiducial tumor motion and external marker motion for liver tumors imaged with 4D-CT*, Int. J. Radiat. Oncol. Biol. Phys **67**, 630 (2007).

*Contact author: C.Graeff@gsi.de

# Supplementary material
## I. Additional setup information

Figure A1(a) shows the setup for the concurrent range measurements with the UCL detector and CNAO detector positioned in sequence to stop the carbon ion beam and helium ion beam respectively. The UCL detector is read out by a set of three cameras, with the distal view being captured through a mirror by a laterally placed camera (Figure A1(b)). The CNAO detector is read out by a single camera placed laterally to the scintillator that captures both the lateral and distal view (via a mirror) simultaneously (Figure A1(c)).

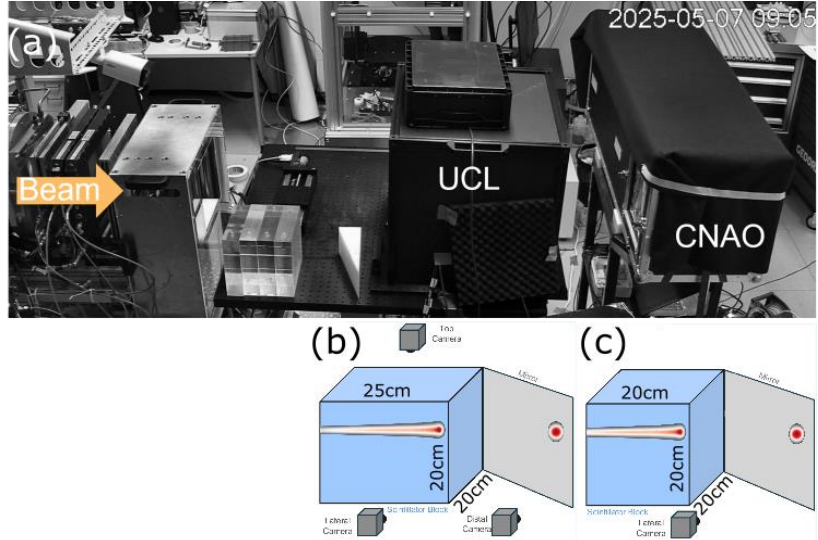

***Figure A1:*** *Setup for concurrent range measurement with the UCL and CNAO detector. (a) Photograph of the setup. The UCL detector is placed first in beam direction, such that the carbon ions fully stop inside the scintillator. The helium beam has sufficient range to cross into the CNAO detector, where it then comes to a stop. (b) A schematic depiction of the UCL detector. The system is read out by three cameras. Two are placed laterally to capture a lateral, and rear view (via a mirror) of the scintillator. One is placed above the scintillator to capture the top view. (c) Schematic view of the CNAO detector. The scintillator is read out by only one camera placed laterally, which captures both the lateral and distal view (via a mirror) simultaneously.*

## II. Film evaluation to derive the carbon and helium beam parameters

To derive the carbon ion beam lateral spread and angular divergence, we evaluated 4 films placed in different distances to the isocenter, as outlined in the main manuscript. The horizontal and vertical integral profile of the scanned films were fitted by a double-Gaussian function to obtain the lateral variance of the beam in either direction. From these measurements it was possible to derive the angular divergence of the beam, by fitting an approximative formulation for the lateral broadening after drift in air to the datapoints. Following Fermi-Eyges' theory [1], the beam lateral variance after a drift in air is given as

$$(1) \quad \sigma_x(z) = \sqrt{\{\sigma_{z_0}^2 + z^2 \sigma_\theta^2 + z \sigma_{x\theta}\}}$$

Where $\sigma_x(z)$ is the variance after drift z, $\sigma_{z_0}^2$ is the variance at isocenter, $\sigma_\theta^2$ is the angular divergence, and $\sigma_{x\theta}$ is the covariance. Assuming the latter to be zero, which is reasonable for a scattered beam, the angular divergence can be derived as a fit to the data. This is shown in Figure A2.

*Contact author: C.Graeff@gsi.de

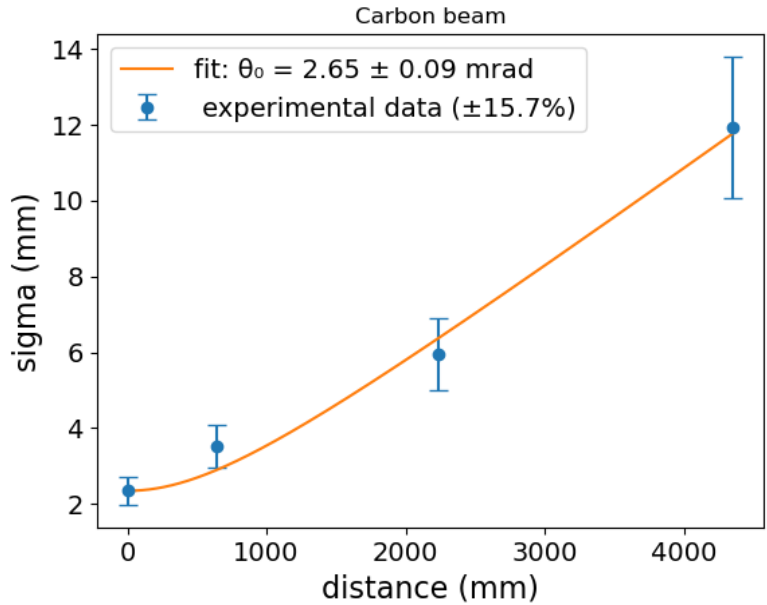

*Figure A2: Standard deviation of a Gaussian fit to the horizontal profile of mixed beam irradiated EBT3 films. The films were positioned without additional material in front of them, such that the carbon ion content of the beam dominates the measurement. The datapoints were fit by a Fermi-Eyges theory-derived analytical formula (Equation 1) to produce the beam angular divergence.*

Based on the carbon ion beam data, Geant4 Monte Carlo simulations were conducted of the helium beam film measurement setup using GATE version 10. Through these measurements, it was verified, that the angular divergence of the carbon beam also accurately reproduced both helium ion film measurements.

### III. Short video of the concurrent range measurement

A short video (in .gif format) was compiled on the concurrent range measurement in the UCL and CNAO detector, which is provided as a separate file. The gif shows a series of frames of 200ms duration each acquired during irradiation of a mixed beam rectangular field of 60 mm (horizontal) times 28 mm (vertical) dimension, at 2x2 mm lateral spot spacing and $5 \times 10^5$ carbon ions per spot, over a 6 cm diameter PMMA sphere placed at isocenter. The helium concentration was set to 'high'. The gif comprises the lateral view captured of the UCL detector, as well as the camera output of the CNAO detector which combines lateral and distal view into one image. The UCL detector image was preprocessed to correct for optical artifacts, such as lens distortion. Several images captured with the UCL detector were summed up to match the measurement frame duration of the CNAO detector (200 ms). The images of both detectors were adjusted such that their lateral dimensions match the physical distance.

### IV. Additional material on carbon ion background removal with deep learning

A deep learning model was developed to remove the background of the carbon ion fragments from the measured helium scintillator signal in silico. The model was based on a U-Net architecture trained on paired, simulated mixed beam and pure helium scintillator images. Figure A3 shows the performance of the model in removing the carbon ion fragment background for one example pencil beam.

*Contact author: C.Graeff@gsi.de

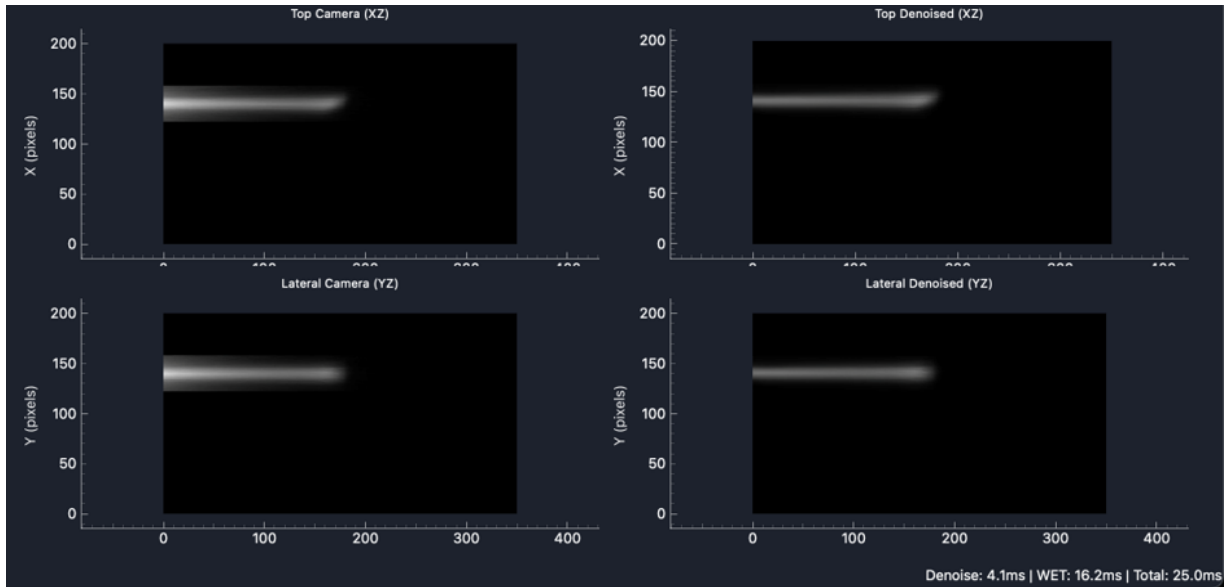

***Figure A3:*** *Left lateral and top scintillator view for a mixed beam crossing a patient. The data is cropped to a 36 times 350 mm² patch around the primary beam, which is also the deep learning model input size. The right column shows the top and lateral view after the U-Net fragment background removal was applied.*

1. Gottschalk, B., *Techniques of Proton Radiotherapy: Transport Theory.* 2012. **arXiv:1204.4470**.

*Contact author: C.Graeff@gsi.de